\documentclass[aps,prb,floatfix,reprint,superscriptaddress]{revtex4-1}
\usepackage{graphicx}
\usepackage{bm,amsmath,amssymb,mathrsfs,dcolumn}
\usepackage[usenames]{color}
\usepackage{subfigure}
\usepackage{ulem}
\usepackage{float}
\usepackage{color}

\usepackage{hyperref}
\hypersetup{colorlinks=true, linkcolor=blue, citecolor=red, urlcolor=blue}

\begin{document}
\title{Current-driven skyrmion motion in granular films}
\author{Xin Gong}
\affiliation{Physics Department, The Hong Kong University of Science and
Technology, Clear Water Bay, Kowloon, Hong Kong}

\author{H. Y. Yuan}
\email[Electronic address: ]{yuanhy@sustech.edu.cn}
\affiliation{Department of Physics, Southern University of Science and Technology,
Shenzhen 518055, Guangdong, China}

\author{X. R. Wang}
\email[Electronic address: ]{phxwan@ust.hk}
\affiliation{Physics Department, The Hong Kong University of Science and
Technology, Clear Water Bay, Kowloon, Hong Kong}
\affiliation{HKUST Shenzhen Research Institute, Shenzhen 518057, China}
\date{\today}

\begin{abstract}
Current-driven skyrmion motion in random granular films is investigated with
interesting findings. For a given current, there exists a critical disorder
strength below which its transverse motion could either be boosted below a
critical damping or be hindered above the critical damping, resulting in
current and disorder dependences of skyrmion Hall angle. The boosting comes
mainly from the random force that is opposite to the driving force (current).
The critical damping depends on the current density and disorder strength.
However, the longitudinal motion of a skyrmion is always hindered by the disorder.
Above the critical disorder strength, skyrmions are pinned.
The disorder-induced random force on a skyrmion can be classified as static
and kinetic ones, similar to the friction force in the Newtonian mechanics.
In the pinning phase, the static (pinning) random force is transverse to the
current density. The kinetic random force is opposite to the skyrmion velocity
when skyrmions are in motion. Furthermore, we provide strong evidences that the
Thiele equation can perfectly describe skyrmion dynamics in granular films.
These findings provide insight to skyrmion motion and should be important for
skyrmiontronics.
\end{abstract}

\maketitle
\section{Introduction}
Magnetic skyrmions have attracted much attention in recent years because
of their potential applications in information storage and processing,
besides their academic interest \cite{Bogdanov2001,Rossler2006,Muhlbauer2009,
Romming,Yu2010,Yu2011,Zhou2014,Yuan2016,Onose,Park,Tian,Heinze,Jiang,Li,news,
Woo2016,Iwasaki2013,Nagaosa,Nagaosa,Fert,Xu,temp_grad,MnSi_anis,size2015,JMMM1,
JMMM2,Lenov2016,Braun,size2016,PdFeIr,BandK,MnSi,Yuan2017, Jiang2017,Kai2017,
Yuan2018,Xiansi,Reich2015,Reich2016,Thiaville2013,Sampaio2013,Lin2013,
koshibas2018,Judge2018,Kim2018,Hoshino2018,Woo2018}.
This potential can only be realized when a good understanding of disorder
effect on skyrmion motion is obtained because defects/inhomogeneity exist
inevitably in all materials. There are a few studies \cite{Woo2016,
Iwasaki2013,Reich2015,Reich2016,Sampaio2013,Lin2013,koshibas2018,Judge2018,Kim2018,
Hoshino2018,Woo2018,fern2018,Juge2019} of skyrmion in disordered systems. Many phenomena were
observed with limited understanding. For example, in the presence of isolated
impurities, micromagnetic simulations suggest that skyrmions sometimes avoid
impurities \cite{Iwasaki2013} and sometimes be trapped by disorders \cite{Sampaio2013}.
There is no well-accepted understanding of these seemingly conflicting results.
The ability of a skyrmion to avoid trapping with detoured trajectory away from
isolated defects was attributed to its topological property \cite{Iwasaki2013}.
This interesting ability leads to the theoretical prediction \cite{Iwasaki2013}
that disorders have little effects on skyrmion driving current density.
Just like their counterparts in a uniform film, skyrmions in a disordered film
should move under a current density as low as $10^5-10^6$ $\mathrm{A/m^2}$, five
orders of magnitudes smaller than that for a domain wall \cite{Iwasaki2013}.
However, the experimental reported driven current is above $10^{10}-10^{11}$
$\mathrm{A/m^2}$ \cite{Woo2016,Jiang2017,Kai2017}, not too far from the typical
magnetic domain wall driving current and far above the community's expectation.
Skyrmions can also perform a random-walk-like motion in disordered system \cite{Nozaki2019}.
All these issues and more need detail analysis and a better understanding.

In this paper, we study the influences of disorders on skyrmion motion.
Our numerical and analytical results show that disorder could boost skyrmion
transverse motion under certain conditions while disorders always hinder
skyrmion longitudinal motion. Thus, the skyrmion Hall angle increases in the
disorder system. The physics behind the boosting (hindering) of the transverse
(longitudinal) motion is mainly from the random forces along the driving force
direction: When the random force is opposite to the driving force, skyrmion
transverse speed increases while the longitudinal speed decrease.
In the opposite case when the random force is along the driving force,
the transverse speed decreases and the longitudinal speed increase.
As a result, the duration time of the skyrmion with a larger (smaller)
transverse (longitudinal) speed is longer than that in the opposite situation.
This explains boosting of the time average transverse skyrmion speed
and the hindering of the average longitudinal speed in disorders.
In comparison, the random force transverse to the current direction decreases
or increases both skyrmion transverse and longitudinal speeds at the same time.
To the first order, the random force in the transverse direction has little
effect on average skyrmion velocity.

The paper is organized as follows.
In Sec. II, we first describe the model and approach adopted in this study.
Sec. III presents our main findings, including three phases, origin of the
boosting of skyrmion transverse motion, static and kinetic random forces, and how
accurate the Thiele equation is in describing  skyrmion motion in granular films.
Interestingly, the average random force on a moving skyrmion is opposite to the
skyrmion velocity. In the discussion, we point out the fundamental differences
between disorder-induced domain wall motion boosting and skyrmion motion boosting,
and show that the physics presented here does not change when the non-adiabatic
torque is present as long as its value is smaller than the damping coefficient.
We will also discuss the spin-orbit torque
driven skyrmion motion.
The conclusion is given in Sec. IV, followed by Acknowledgements.

\section{Model and methodology}
We consider a perpendicularly-magnetized random granular film that is
constructed by a Voronoi tessellation, as shown in Fig. \ref{fig1}(a).
Magnetic anisotropy $K$ in each grain is randomly distributed around
$K_0=8.0\times 10^5~\mathrm{J/m^3}$ either with a Gaussian distribution
of deviation $\Delta K$ or in a window of $(K_0-\Delta K, K_0+\Delta K)$
The granular film with Gaussian distribution is
assumed below if it is not specified.

The material parameters are chosen in such a way that supports a stable isolated
skyrmion \cite{Xiansi}. Initially, a skyrmion is located in the center (also the
origin of the $xyz$-coordinate).
Spin dynamics is governed by the Landau-Lifshitz-Gilbert (LLG) equation,
\begin{equation}
\frac{\partial \vec m}{\partial t} =-\gamma\vec m \times \vec H_{\rm eff} +
\alpha \vec m \times \frac{\partial \vec m}{\partial t} + \vec \tau,
\label{llg}
\end{equation}
where $\vec m$, $\gamma$, $\alpha$ are respectively the unit vector of the
magnetization, gyromagnetic ratio, and the Gilbert damping.
$\vec H_{\rm eff}=2A \nabla^2\vec m+2Km_z\hat z
+\vec H_d+\vec H_{\rm DM}$ is the effective field including the exchange
field characterized by the exchange stiffness $A$, crystalline anisotropy field
along the $z$-direction, dipolar field $\vec H_d$, and the Dzyaloshinskii-Moriya
interaction (DMI) field $\vec H_{\rm DM}$ characterized by DMI coefficient $D$.
In this study, we consider the interfacial DMI with DMI energy density of
$D[(\hat z\cdot \vec m)\nabla \cdot \vec m-(m\cdot \nabla)(\hat z \cdot \vec m)]$.
$\vec \tau= -(\vec u\cdot \nabla )\vec m +\beta\vec m \times (\vec u\cdot\nabla )
\vec m$ is the spin transfer torque consisting of a damping-like torque and a
field-like torque \cite{Zhang2004,Thiaville2005}, where $\beta$ is a dimensionless
coefficient measuring the strength of the field-like torque and $\vec u=\vec JP
\mu_B/(eM_s)$ is of a quantity of dimension of speed measuring amount of equivalent
magnetic moments supplied from spin polarised current to the magnet per unit time.
Here $J,P,\mu_B, e$, and $M_s$ are respectively current density, current polarization,
the Bohr magneton, the electron charge, and the saturation magnetization.
To study the skyrmion dynamics, we use the Mumax3 package \cite{mumax3} to numerically
solve the LLG equation on the granular film of size of $1024$ nm$\times 512$ nm
$\times 1$ nm. The mesh size is $1$ nm$\times 1$ nm $\times 1$ nm.
This choice of mesh size is tested through the non-change in simulation results when 
a smaller mesh size is used (See Appendix for the results of smaller mesh size).
The average grain size is of 5 nm, and the model parameters are
$A=15\times 10^{-12}~\rm{J/m},\ D=0.003~\rm{J/m^2},\ M_s=5.8\times 10^5 ~\rm{A/m}$
in this study.

To understand the skyrmion dynamics, we derive the Thiele equation in disorders
by considering $\partial_i \vec m \cdot (\vec m \times$ Eq. (\ref{llg})).
Under the rigid-body assumption and after some algebras, we obtain following
equation for skyrmion-velocity $\vec v$
\begin{equation}
\vec G\times(\vec v-\vec u)+\tensor D\cdot (\alpha\vec v-\beta\vec u)+
\frac{\gamma}{M_s d}\nabla E=0,
\label{gt}
\end{equation}
where $d$ is the film thickness, $\vec G=G\hat z=4\pi Q\hat z$ is the skyrmion
gyrovector proportional to the skyrmion number $Q$ \cite{Xiansi}, and the tensor
$D_{ij}=\int \partial_i\vec m \cdot \partial_j \vec m dS$ is the dissipation dydic.
For symmetric skyrmion structures, $D_{ij}=D\delta_{ij}=\frac{1}{2}(R/w+w/R)\delta_{ij}$
where $R$ and $w$ are respectively skyrmion size and skyrmion wall width \cite{Xiansi}.
$E(\vec R)=\int\int \{A(\nabla\vec m)^2+D[(\hat z\cdot \vec m)\nabla \cdot \vec m-
(m\cdot \nabla)(\hat z \cdot \vec m)]-K(\vec x, \vec y)m_z^2 \}dxdy$ is the total
energy of a skyrmion centred at position $\vec R$. In a homogeneous film, the total
energy of the skyrmion does not depend on the skyrmion position ($\nabla E=0$) due to
the translational symmetry of the system.
Eq. \eqref{gt} is the original equation derived by Thiele. \cite{thiele}
In a granular film, the translational
symmetry breaks so that $\nabla E \neq 0$. $\vec F \equiv -\frac{\gamma}{M_s d}
\nabla E$ is a random force in all directions when a skyrmion is in motion.
At the moment, we set $\beta=0$ and $\vec F=(F_x,F_y)$ and Eq. \eqref{gt} becomes,
\begin{subequations}
\begin{align}
&Gv_x+\alpha D v_y=Gu+F_y \\
&Gv_y-\alpha D v_x=-F_x
\end{align}
\end{subequations}
Without losing generality, we set $Q=1$ and assume $u>0$. The solution of the above
skyrmion dynamical equations is $v_x=[G^2u+GF_y+\alpha D F_x]/[G^2+(\alpha D)^2]$,
$v_y=[\alpha DGu+\alpha D F_y-GF_x]/[G^2+(\alpha D)^2]$. In a homogeneous film, we
have $F_x=F_y=0$ and $v_y>0$ and $0<v_x<u$ since $\alpha D>0$. Furthermore, the
skyrmion Hall angle does not depend on $u$ (current).
For a negative (positive) $F_x$ and $F_y=0$, $v_y$ increases (decreases) and
$v_x$ decreases (increases), skyrmion Hall angle depends on both disorders and $u$.
In contrast, $v_x$ and $v_y$ vary with $F_y$ differently when $F_x=0$:
$v_x$ and $v_y$ increase (decrease) simultaneously for $F_y>0$ ($F_y<0$).
These dependences of $v_x$ and $v_y$ are important for us to understand the
boosting of skyrmion transverse discussed below.

\section{Result}
\subsection{Three Phases}

\begin{figure}
\centering
\includegraphics[width=1.0\columnwidth]{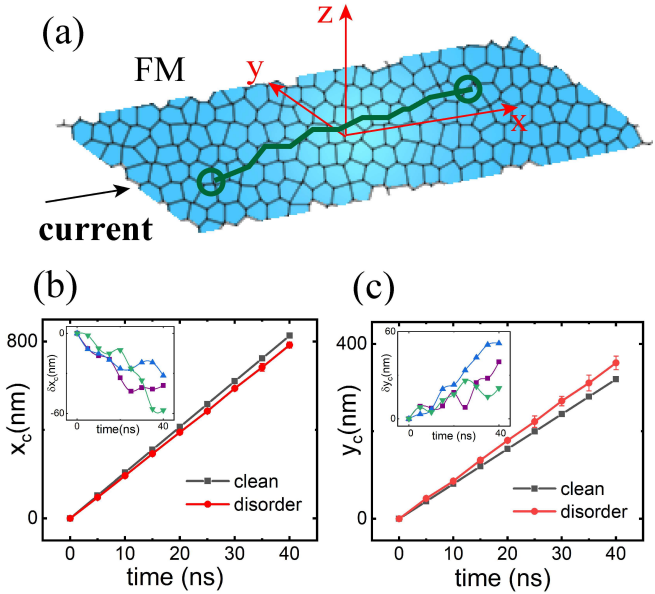}
\caption{(color online)(a) Schematic illustration of a skyrmion in a
granular chiral magnetic film with randomly distributed anisotropy.
(b) and (c) Time evolution of 10 ensemble averaged skyrmion position $\vec R=(x_c,\
y_c)$ under an electric current for a granular film of $\Delta K=3\% K_0$ (red dots).
For a comparison, black squares are for a homogeneous film. The inset shows the
position difference between the homogeneous and three typical granular films.
The three curves are for three different realizations.
The model parameters are $\alpha=0.3,\ \beta =0,\ J=6\times10^{11} \rm{A/m^2}$.}
\label{fig1}
\end{figure}

\begin{figure}
\centering
\includegraphics[width=0.9\columnwidth]{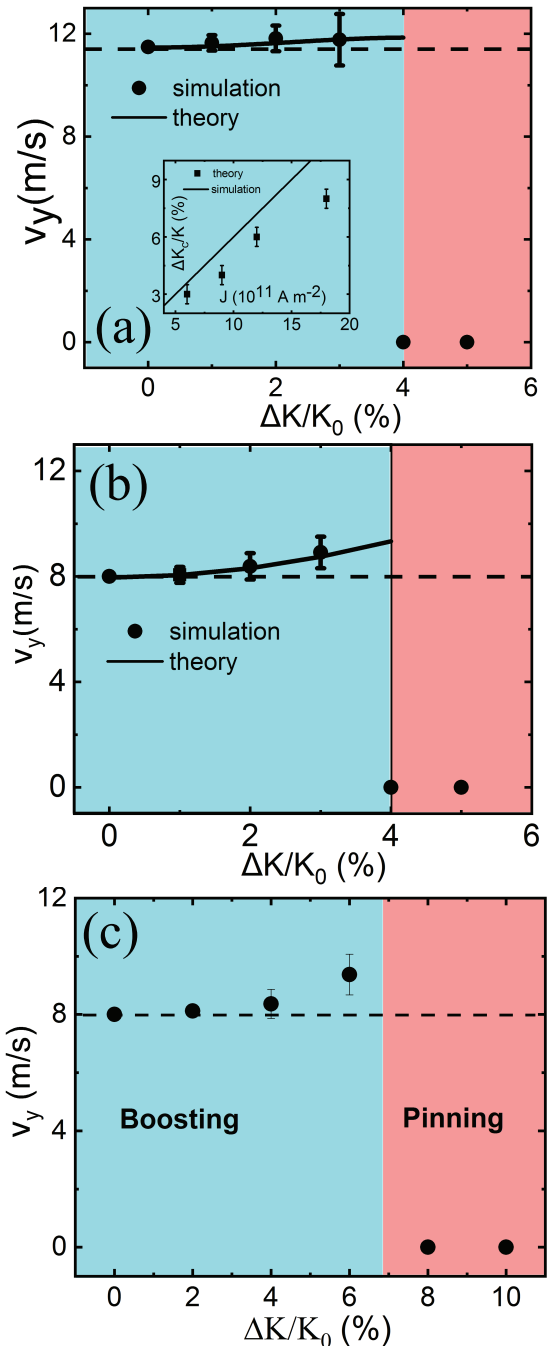}
\caption{(color online) Disorder strength dependence of skyrmion transverse velocity
for $J = 6\times10^{11} \rm{A/m^2}$. (a) and (b) are
for Gaussian distribution of magnetic anisotropy with $\alpha=0.6$ (a) and $\alpha=0.3$ (b). (c) is
the same plot for the uniform distribution of magnetic anisotropy with $\alpha=0.3$.
The horizontal dashed line is $v_y$ of the homogeneous film while dots are simulation
results and the solid line is the solution of Eqs. (2) and (4) for $\beta=0$ with
a fitting parameter $b$ defined in random force $f$.
The cyan and red colors denote the boosting phase and pinning phase, respectively.
The inset shows the critical disorder strength as a function of the current density
for $\alpha=0.3, \beta=0$, the black squares are numerical results. }
\label{fig2}
\end{figure}

\begin{figure}
\centering
\includegraphics[width=1.1\columnwidth]{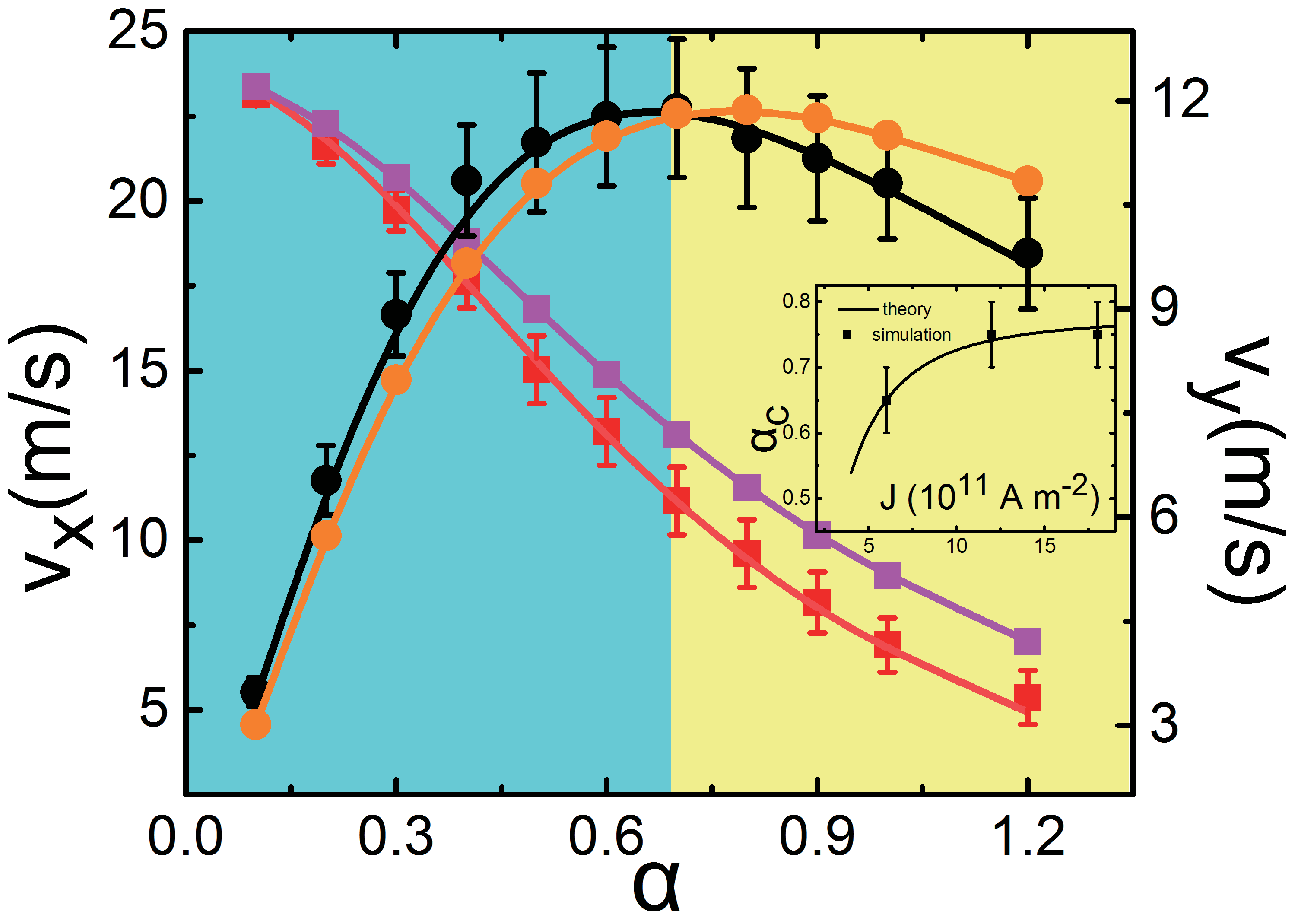}
\caption{(color online) The $\alpha-$dependence of longitudinal (squares) and transverse
(dots) velocities for $J =6\times 10^{11} \rm{A/m^2}$ and $\Delta K = 3\%K_0$.
The orange and magenta colors denote the velocities in the homogeneous film while
the black and red colors are for the granular film. The lines are the solutions
of Eqs. (2) and (4) with $b$ in $f$ (for granular film) as a fitting parameter.
The cyan and yellow colors denote the boosting phase and hindering phase, respectively.
The inset shows the current density dependence of critical damping.
Other parameters are $\beta=0$ and those specified in the model.}
\label{fig3}
\end{figure}

Figures \ref{fig1}(b) and \ref{fig1}(c) plot skyrmion positions $\vec R=(x_c,y_c)$ as a
function of time for a homogeneous film (black squares) and for a granular film (red dots)
of $\Delta K=3\% K_0$ for $\alpha=0.3,\ \beta =0,\ J = 6\times10^{11} \rm{A/m^2}$. $x_c$
and $y_c$ for the granular film are the ensemble average over 10 independent realizations.
The inset shows the position differences of three typical granular films and the
homogeneous film. The trajectory of skyrmion motion in the homogeneous film
is a perfectly straight line with constant velocity, while it wiggles around a
straight line in each realization of random granular film as shown in the insets.
The ensemble average of $x_c(t)$ and $y_c(t)$ are perfectly linear so that average
skyrmion velocity $\vec v$ is a good description of skyrmion motion.
The average longitudinal and transverse skyrmion velocities are $v_x=19.7$ m/s and
$v_y=8.9$m/s for random granular films, $v_x=20.7$m/s and $v_y=8.0$m/s for the homogeneous
film, which agree perfectly with solution of Eq. (2) with $\nabla E =0$ (black lines).
Interestingly, the transverse motion is boosted by the disorder while the longitudinal
motion is hindered.

One can compute the average skyrmion velocity and its statistical errors from
different realizations and for different disorder strengths and Gilbert damping.
Fig. \ref{fig2} shows how averaged $v_y$ and its error bar vary with the
disorder strength $\Delta K/K_0$ (in percentage) for current density
$J=6 \times 10^{11} ~\mathrm{A/m^2}$ and $\alpha=0.6$ (a) and $\alpha=0.3$ (b).
Points of zero velocity correspond to the skyrmion pinning, similar to the
magnetic domain wall pinning by disorders or notches \cite{yuan1,yuan2,yuan3}.
The critical disorder strength, above which all skyrmion are pinned, depends on the
current density $J$ (proportional to $u$). In cyan region, the average transverse
velocity in the granular film is larger than that in the homogeneous film, and system
is in the boosting phase. As disorder strength $\Delta K/K_0$ increases, the error bars
increase and the average transverse velocities in disorder systems increase parabolic
for $\alpha=0.3$. For a larger disorder strength $\Delta K/K_0 > 4\%$, the skyrmions are in pinning
phase (red region). The inset shows a linearly $J$-dependence of the critical disorder
$\Delta K/K_0$, like the effect of the friction force in a Newtonian mechanics that
one needs a larger driven force to maintain the motion of a body on a rougher surface. 
The physics does not depend on whether the distribution function of random $K$ is Gaussian 
or uniform in a window. As shown in Fig. \ref{fig2}(c) for $K$ uniformly distributed 
in a window of $[K_0-\Delta K, K_0+\Delta K]$, one can clearly see both boosting and pinning. 
The only difference is at qualitative level.

Fig. \ref{fig3} shows how $v_x$ (squares) and $v_y$ (dots) change with the Gilbert
damping coefficient $\alpha$ for $J =6\times 10^{11} \mathrm{A/m^2}$, $\beta=0$,
and $\Delta K=3\%K_0$. Different from $v_x$ that monotonically decreases with $\alpha$
(red), $v_y$ (black) increases first and then decreases with $\alpha$ as shown in Fig.
\ref{fig3}. To see boosting and hindering effect, we have also plotted $v_x$ and $v_y$
for the homogeneous film (orange and magenta respectively) for the same model parameters.
Obviously, skyrmion transverse motion is boosted (hindered) by the disorder for
$\alpha<0.7$ ($\alpha>0.7$). Interestingly, the critical damping that separates
boosting from hindering coincides with the peak position of $v_y$. The longitudinal
motion is, however, always hindered by the disorder in our simulations. The value of
the critical damping coefficient depends on the current density as shown in the inset.

In summary, three phases are identified: Pinning phase above the critical disorder
strength; boosting of skyrmion transverse motion below the critical disorder
strength and below a critical damping coefficient; hindering of skyrmion transverse
motion below the critical disorder strength and above the critical damping.
Both critical disorder strength and critical damping coefficient depend on the
applied current density and other model parameters. Numerically, critical damping
$\alpha_c$ at disorder strength $\Delta K_c/K_0 =3\%$ seems coincide with the peak
position of $v_y$.

\begin{figure}
\centering
\includegraphics[width=1.0\columnwidth]{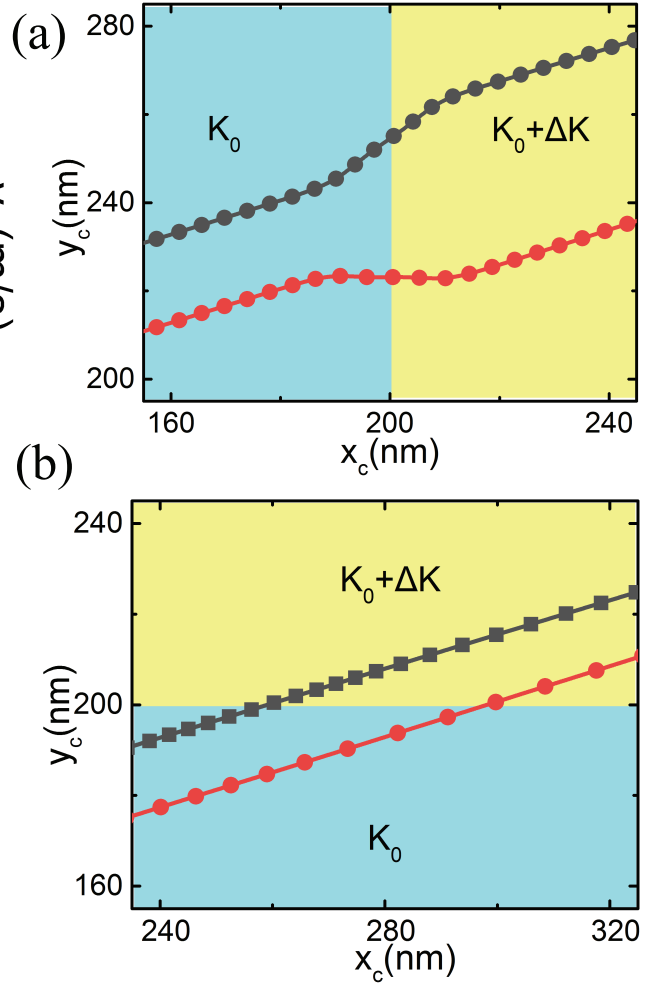}
\caption{(color online) Model parameters in this figure are $J =6\times 10^{11}
\rm{A/m^2}$, $\alpha=0.3, \beta=0$ and $K_0=8.0\times 10^5~\rm{J/m^3}$, as well
as those specified in the model. (a) A y-aligned boundary. The anisotropy is $K_0$
on the left of the boundary, and $K_0+\Delta K$ on the right of the boundary.
The lines indicate the skyrmion trajectories when it cross the boundary from the left.
Black lines for $\Delta K=3\% K_0$ ($F_x<0$) and red line for $\Delta K=-3\% K_0$($F_x>0$).
(b)A x-aligned boundary. The anisotropy is $K_0+\Delta K$ above the boundary, and
$K_0$ below the boundary. The lines indicate the skyrmion trajectories when it cross
the boundary from the bottom. Black line for $\Delta K=3\% K_0$ ($F_y<0$) and red line
for $\Delta K=-3\% K_0$ ($F_y>0$).}
\label{fig4}
\end{figure}

\begin{figure}
\centering
\includegraphics[width=1.0\columnwidth]{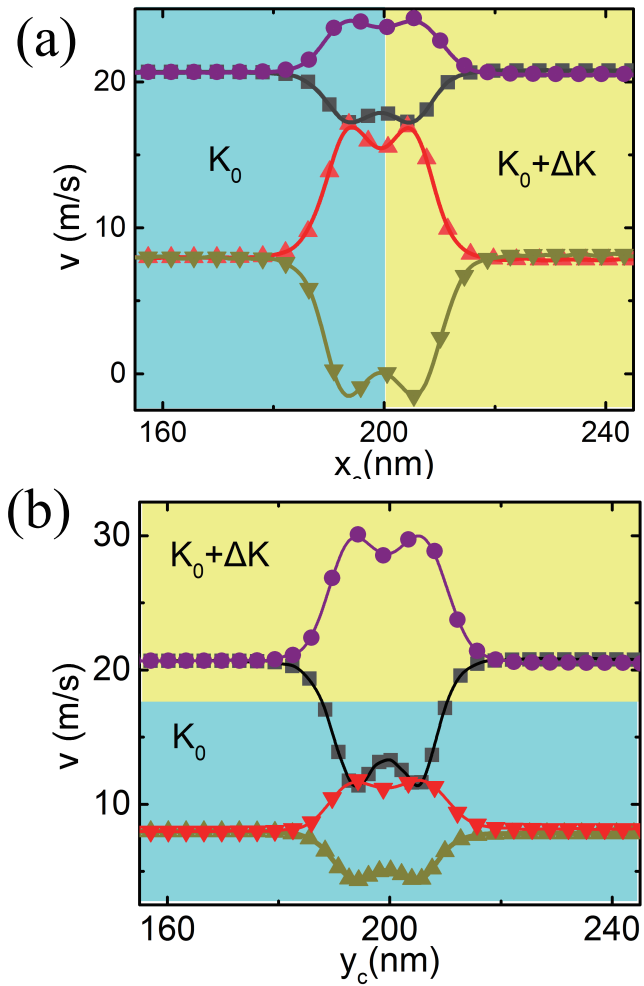}
\caption{(color online) Model parameters in this figure are $J =6\times 10^{11}
\rm{A/m^2}$, $\alpha=0.3, \beta=0$ and $K_0=8.0\times 10^5~\rm{J/m^3}$, as well
as those specified in the model. (a) $v_x$ and $v_y$ at different position for
a y-aligned boundary. Squares and up-triangles are of simulations results of
$v_y$ and $v_x$, respectively, for $\Delta K=3\% K_0$ while dots and
down-triangles are of $v_y$ and $v_x$ for $\Delta K=3-\% K_0$
(b) $v_x$ and $v_y$ at different position for a x-aligned boundary.
Squares and up-triangle are of simulations results of $v_x$ and $v_y$, respectively,
for $\Delta K=3\% K_0$ while dots and down-triangles are for $\Delta K=-3\% K_0$.
All lines in the two figures are numerical solutions of Eq. \eqref{gt} in which 
instantaneous $E$ and $\tensor D$ obtained from skyrmion
structure are used.}
\label{fig5}
\end{figure}

\subsection{Origins of Boosting}

To understand the origins of the boosting of skyrmion transverse motion, we consider
how a skyrmion cross a y-aligned boundary (Fig. \ref{fig4}(a)) and a x-aligned boundary
(Fig. \ref{fig4}(b)) that separates two otherwise homogeneous magnetic films.
For a y-aligned boundary, the boundary force on a rightward moving skyrmion is along the
positive x-direction, $F_x>0$, when the magnetic anisotropy of the film on the left is
larger than $K$ on the right (all other model parameters are the same as specified early).
Fig. \ref{fig5}(a) shows clearly that the average transverse velocity (down-triangles) becomes
smaller while the longitudinal velocity (dots) is larger near the boundary.
The skyrmion trajectory (red dots and line) is defected towards x-direction near the
boundary as indicated by the red line in Fig. \ref{fig4}(a).

If $K$ on the left is smaller than $K$ on the right, the boundary
force is negative, $F_x<0$, and $v_y$ (up-triangles) is larger near the boundary while
$v_x$ (squares) becomes smaller near the boundary shown in Fig. \ref{fig5}(a).
Thus, the skyrmion trajectory is defected towards y-direction near the boundary as
indicated by the black dots and line in Fig. \ref{fig4}(a). This feature was also
observed by others \cite{Manchon}, and was termed as gliding motion of skyrmions.
However, its true origin was not sufficiently revealed. It should be pointed out
that all the lines in Fig. \ref{fig5} are the numerical solutions of Eq. \eqref{gt}
in which $E$ and $\tensor D$ are numerically computed from the spin structures in
simulations that vary with time. The perfect agreement between micro-magnetic
simulations and numerical solutions of Eq. (2) demonstrates excellent approximation
of the Thiele equation although the rigid-body assumption is obviously invalid for 
a skyrmion cross a boundary. For a randomly distributed disorders, the sizes of regions
with $F_x>0$ and $F_x<0$ should be the same so that a skyrmion spend more time stay
in $F_x<0$ regions than that in $F_x>0$ ones. As a result, the time average skyrmion
transverse velocity is boosted while the longitude velocity is hindered.

In contrast, when a skyrmion crosses a x-aligned boundary (\ref{fig4}(b)), both $v_x$
(dots and squares) and $v_y$ (down-triangles and up-triangles) increase near the
boundary for $F_y>0$ and decrease near the boundary for $F_y<0$, as shown in Fig. \ref{fig5}(b).
Again, the lines in the figures are the numerical solutions of Eq. (2).
Different from the effect of $F_x$, random force in the transverse direction has no
much effect on the skyrmion hall angle because it increases (decreases) longitudinal
and transverse simultaneously so that skyrmion trajectories show negligible deflection
when skyrmions cross the boundary as shown by the red and black lines in Fig. \ref{fig4}(b).
From Fig. \ref{fig5}(b), a velocities difference of about $0.1 m/s$ in the left and right
domain far away from the boundary is observed that is much smaller than the boundary effect.
Thus, influence of the boundary force dominant and small variation of $\tensor D$ in
different domains will be neglected in the following analysis. Follow the same
analysis as the case of $F_x$, we found weak hindering of both $v_x$ and $v_y$.
In summary, both y-aligned and x-aligned boundary effect lead to the hindering of longitude
motion. However, y-aligned boundary effect is the main cause of boosting of transverse motion.

\subsection{Random Force at Pinning and in Motion}

\begin{figure}
\centering
\includegraphics[width=1.0\columnwidth]{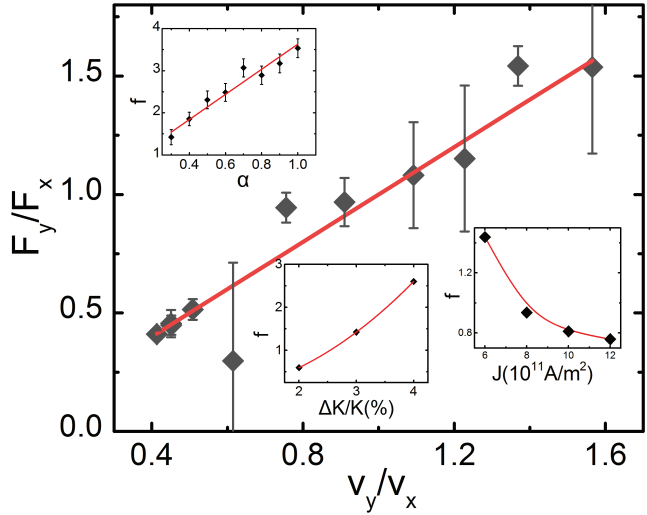}
\caption{(color online) $F_y/F_x$ vs. $v_y/v_x$ for various $J$, $\Delta K$, and
$\alpha$. All other model parameters are specified in the section of model.
Filled diamonds are the numerical data, and the black line is for $y=x$.
All ($F_y/F_x$, $v_y/v_x$) fall around the line. Insets show the $f$ dependence on
$J$, $\Delta K$, and $\alpha$. }
\label{fig6}
\end{figure}

In order to have a better understanding of three phases, we would like to consider
the random force defined in Eqs. \eqref{gt} and (3). For simplicity, we consider
only the case of $\beta=0$. When a skyrmion is at rest, then random force should
balance the driven force from current $\vec u$, i.e. $\vec F=\vec G \times \vec u$.
Thus the static random (pinning) force must be transverse to the current direction.
To study the random force on a skyrmion when it moves in a random potential landscape,
we would like to numerically verify our conjecture that, like the friction force of an
object moving on a surface that is always opposite to its velocity in Newtonian mechanics,
average random force is also opposite to the skyrmion motion. Since Thiele equation is an
excellent description of skyrmion dynamics in domains and crossing domain boundaries,
we can substitute the instantaneous skyrmion velocity obtained from the micromagnetic
simulations into Eq. (3) to obtained the numerical random force for a given system at
each moment. Obviously, this is a stochastic quantity that varies from time to time.
In terms of skyrmion motion, the meaningful quantity is the time averaged random force.
Below, all $F_x$ and $F_y$ are the time averaged values for each given system.
One can obtain different pairs of ($F_y/F_x$, $v_y/v_x$) by using different $J$,
$\alpha$, and $\Delta K$, and they should fall on line $y=x$ if $\vec F \parallel -\vec v$.
Fig. \ref{fig6} plot points of ($F_y/F_x$, $v_y/v_x$) in the $F_y/F_x-(v_y/v_x)$ plane.
It is clear that all points of ($F_y/F_x$, $v_y/v_x$) lie indeed around line $y=x$, a
strong numerical evidence that the direction of $\vec F$ is opposite to skyrmion velocity,
\begin{equation}
  \vec F = - f\hat v,
\label{pin}
\end{equation}
where $\hat v$ is the unit direction of skyrmion velocity.

$f$ depends on the driven current, disorder strength as well as the skyrmion structure.
From Eq. (3), the average velocity $v_x = \frac{G^2 u-(\alpha D F_x)^2/G^2u} {G^2+(\alpha D)^2}$
and $v_y = \frac{\alpha D G u+\alpha D F_x^2/G u}{G^2+(\alpha D)^2}$. Assume the regions of
$F_x<0$ and $F_x>0$ are the same, the time average random force should be $F=F_x(1/v_{x1}
-1/v_{x2})/(1/v_{x1}+1/v_{x2})= -F_x(v_{x1}-v_{x2})/(v_{x1}+v_{x2})\simeq -(\alpha D F_x^2)
/G^2u$, where $v_{x2}$ ($v_{x1}$) is of the velocity under random force $-F_x$ ($F_x$).
This suggests that $f$ takes the form of $f=b \alpha D \Delta K^2/ J=b\alpha D \delta/u$,
where $\delta\equiv \Delta K^2 P\mu_B/(eM_s)$ measures disorder strength and has a
dimensionality of velocity. $b$ is a dimensionless numerical factor of order of 1.
To test this reasoning, we numerically plot $f$ against $\alpha$, $\Delta K$, and $J$ in
the insets of Fig. \ref{fig6}. As shown in the insets, $f$ is proportional to $\alpha$,
$\Delta K^2$, but inversely proportional to $J$, as conjectured.

We can test how good of this $f$ is by substituting Eq. \eqref{pin} into Eq. (3), and
solve for $\vec v$ as a function of $\Delta K/K $ and $\alpha D$ by treating $b$ as the
only fitting parameter. For the small pinning strength ($\Delta K/K_0 \leq 3\%$), we
carried out the calculation for the model parameters used for simulations in Figs.
\ref{fig2} and \ref{fig3}, and the lines are the theoretical results.
Almost perfect agreement between the simulation results and numerical solution of Eq. (3)
demonstrates not only high accuracy of the Thiele equation, but also the excellent
approximation of $f$.

\subsection{Phase Diagram}

By substituting $\vec F$ in Eq. \eqref{pin} into the generalized Thiele equation (\ref{gt}),
we can solve the equation for skyrmion velocity,
\begin{subequations}
\begin{align}
&v_x=\frac{G^2}{G^2+(\alpha D+f/v)^2}u \label{vx}\\
&v_y=\frac{G(\alpha D+f/v)}{G^2+(\alpha D+f/v)^2}u \label{vy}\\
&v=\frac{\sqrt{(u^2-f^2)(G^2+(\alpha D)^2)+\alpha^2D^2f^2}-\alpha Df}{{G^2+(\alpha D)^2}}
\end{align}
\label{vxvy}
\end{subequations}
where $v\equiv|\mathbf{v}|=\sqrt{v_x^2+v_y^2}$.
Clearly, Eq. (\ref{vx}) shows that the longitudinal velocity ($v_x$) is always
hindered by the random force $f$. Eq. (\ref{vy}) suggests the existence of a
maximum $v_y$ at $\alpha D+f/v = G$, leading to a critical damping $\alpha_c= G/[(1+
\frac{b\Delta K^2}{Jv})D]$ that separate boosting phase from the hindering phase.
To pin a skyrmion, the random force proportional to $\Delta K$ must balance
the driven force of magnitude $Gu$. Thus the critical pinning disorder strength should
be of $\Delta K_c=cGu$, or $\Delta K_c/J=c P\mu_B/(eM_s)$, where $c$ is a factor that
depends on the skyrmion size and structure.

\begin{figure}
\centering
\includegraphics[width=1.0\columnwidth]{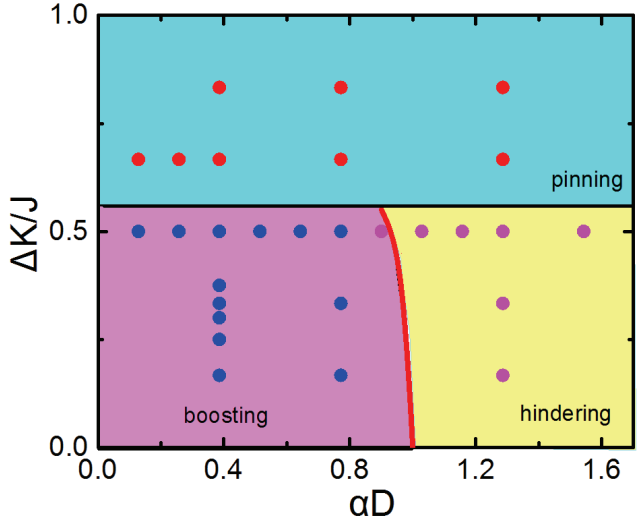}
\caption{(color online) Three phases in $\Delta K/J$-$\alpha D$ plane. Dots are simulations
results for pinning (red), boosting (blue) and hindering (pink) phases. The two boundary
lines separating three phases. The black line are pinning-unpinning boundary while the
red line are boosting-hindering boundary}
\label{fig7}
\end{figure}

As shown in Fig. \ref{fig7} in the plane of $\Delta K_c/J$-($\alpha D$), $\Delta K_c/J=c
P\mu_B/(eM_s)$ (black line with an optimal $c$) separates the pinning phase from the unpinned
phase, and $\alpha_c= G/[(1+\frac{b\Delta K^2}{Jv})D]$ (red line) further separate the
boosting phase from the hindering phase. Since the boundaries are obtained from the average
random force which is smaller than the maximal possible force in the granular film, one
should expect the critical disorder strength under-estimate the pinning since it the maximal
possible force from the random potential landscape that is relevant for the pinning.
Indeed, as one can see the true simulations results (insets in Fig. \ref{fig2}), pinning
occurs at a disorder strength below the theoretical prediction.

\begin{figure}
\centering
\centering
\includegraphics[width=1.0\columnwidth]{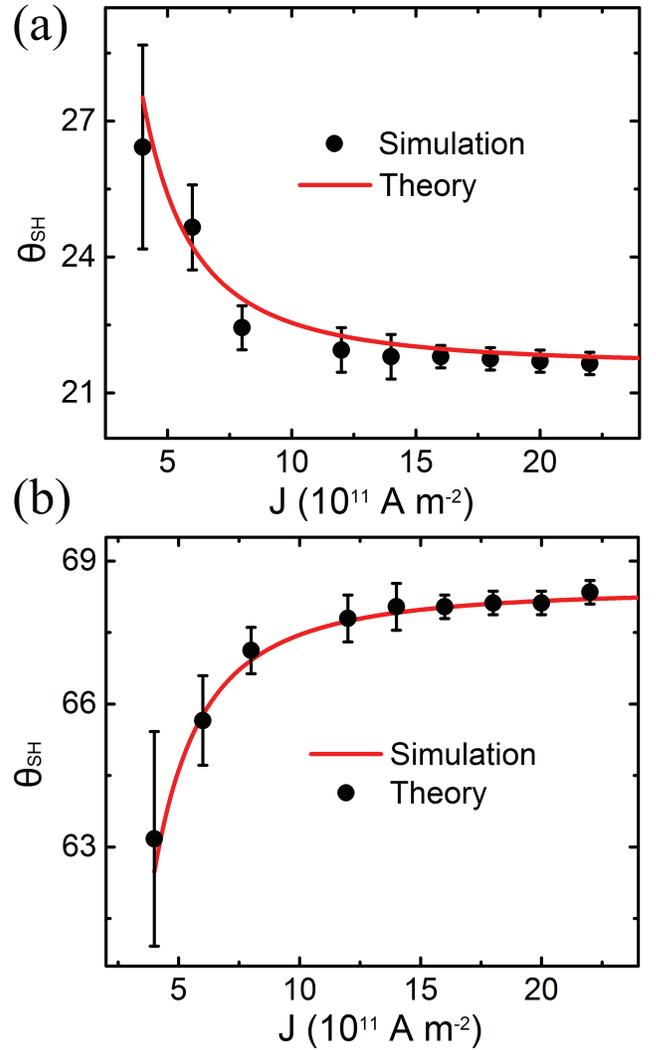}
\caption{(color online) The current density dependence of skyrmion Hall angle $\theta_{\mathrm{SH}}$
for the STT driven (a) and SOT driven (b) cases.
Red lines are Eq. (\ref{stt-hallangle}) and
Eq. (\ref{sot-hallangle}), black circles are numerical results.
Other parameters are $\Delta K=3\% K_0, \beta=0$}
\label{fig8}
\end{figure}

From Eq. (5), one can also obtain the skyrmion Hall angle
\begin{equation}
\theta_{\mathrm{SH}} = \tan^{-1} \left ( \alpha D + \frac{\delta}{u^2}\alpha D\sqrt{1+\alpha^2D^2} \right ).
\label{stt-hallangle}
\end{equation}
It predicts that the Hall angle decrease gradually with current density and approaches a constant value $\alpha D$ that
is the hall angle for the homogeneous film. As shown in Fig. \ref{fig8}(a), this formula (red line)
could well capture the trend of numerical results (dots). The discrepancy may come from the current
density dependence of the pinning strength $\delta$ and the deformation of moving skyrmion that
is not included in our model.

\section {Discussions and conclusions}

\begin{figure}
\centering
\includegraphics[width=0.9\columnwidth]{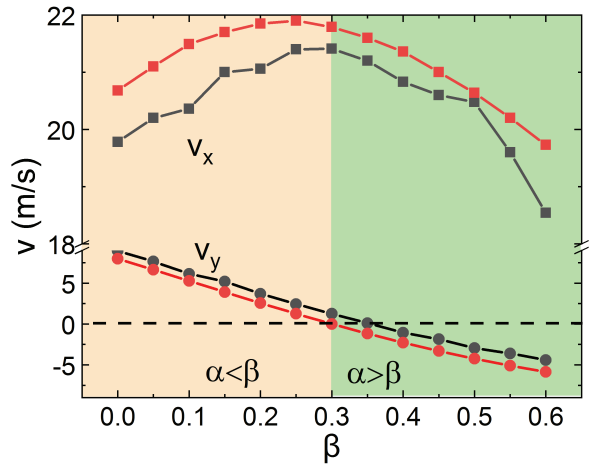}
\caption{(color online) The $\beta-$dependence of longitudinal (squares) and transverse
(dots) velocities for $\alpha=0.3$, $J =6\times 10^{11} \rm{A/m^2}$ and $\Delta K = 3\%K_0$.
The red colors denote the velocities in the homogeneous film while
the black colors are for the granular film. }
\label{fig9}
\end{figure}

It was known that disorders can also boost magnetic domain wall propagation \cite{yuan2,yuan3}.
The boosting there is related to the generation of anti-vortices that can both help domain
wall depining \cite{yuan2,yuan3} and exert an extra driving force through Magnus effect.
Thus, it is very different from the physics of disorder-boosted transverse skyrmion motion
that come from the random force opposite to the current direction. So far, all simulations are
for $\beta=0$. In realistic system, non-adiabatic torque $\beta$ should not be zero in general
although its value is believed to be small. Thus, it is nature to ask whether the physics will
be different when non-zero $\beta$ is considered. To address this issue, we have also carried out
the same simulation for non-zero $\beta$ and the results are shown in Fig. \ref{fig9}. 
The longitudinal velocity is always hindered by the disorders regardless of value of $\beta$.
Note that the transverse velocity reverses its sign at $\beta=\alpha$. Thus, the transverse 
skyrmion motion is always boosted for $\beta<\alpha$ while the boosting is absent when $\beta>\alpha$.
Thus, the main results reported here is valid only when $\beta<\alpha$.

Although our simulations and
theory focus on the current-driven skyrmion motion thorough spin transfer torque (STT), all the
physics is still valid for skyrmion motion through spin-orbital torque, and the corresponding
Thiele equation is \cite{Jiang2017,KLwang},
\begin{equation}
\vec G \times \vec v +\alpha \tensor D \cdot \vec v -\tensor B \cdot \vec u'- \vec F=0,
\label{sotthiele}
\end{equation}
where the element of $\tensor B$ is $B_{ij}=\iint\int (\partial_i m_zm_j-m_z\partial_i m_j)dxdy $.
and $u' = (\gamma\hbar\theta_{\rm {SH}})/(2eM_sd)$
The Hall angle from the the new Thiele equation becomes
\begin{equation}
\theta_{\rm{SH}}=\pi/2-\tan^{-1}\left( \alpha D+\frac{\delta}{u^2}\alpha D\sqrt{1+\alpha^2D^2} \right ).
\label{sot-hallangle}
\end{equation}
The driving current dependence of skyrmion Hall angle has been
observed in both experiments and micromagnetic simulations \cite{Jiang2017, Kai2017, Reich2015}.
But, to the best of our knowledge, analytical formula like Eqs. \eqref{stt-hallangle} and
\eqref{sot-hallangle} for the skyrmion Hall
angle were not known.

Figure \ref{fig8}(b) is of the comparison of our theory with the micromagnetic simulations (dots).
Different from the STT-driven case, the Hall angle increases gradually to the value of the homogeneous
film with the increase of current density $J$. Our theory (red line) captures well this trend.

How to manipulate and control skyrmion Hall angle is an important issue in device applications
because non-zero Hall angle tends to pushes skyrmions to sample edges, leading to skyrmion annihilation.
Even though the Hall effect itself comes from skyrmion topological structure that seems have
nothing to do with disorders, two independent experiments showed recently that the skyrmion
Hall angle ($\theta_{\mathrm{SH}} \equiv\arctan (v_y/v_x)$ first increases with current
density and then saturates at a sufficiently large value \cite{Kai2017,Jiang2017}.
So far, a good understanding of the observed behavior of Spin Hall angle is still lacking
although there are simulations \cite{Reich2015,Woo2018} showing the saturation behavior.
So far, most theoretical studies considered only isolated defects. Although isolated defects
are important in real systems, systems with continuous random grain boundaries may be more
relevant for amorphous and poly-crystal films \cite{Woo2018,Kim2018}.

It should be useful to compare our findings with recent works.
Ref. \cite{Reich2016} treats skyrmions as point-particles and artificially treats disorder
effect as a harmonic potential on skyrmions. This assumption is not well justified and its
prediction cannot compare against micromagnetic simulations and experiments. In contrast, our
random force expression is well justified and compared with simulations.
The resulting results for the skyrmion Hall angle describe well both STT and SOT driving skyrmions
that have opposite current dependence. Ref. \cite{fern2018} is on the interaction of skyrmions 
with atomic defects in PdFe/Ir and it is different from what we have done here.
Also, the first-principles calculations are more like experiments that do not automatically
provide the physics. As an example, the reference did not obtain the random force reported here.
Moreover, the dynamics of skyrmion as well as the skyrmion Hall effect were not studied
in the mentioned paper.

In conclusion, we have investigated the skyrmion motion in inhomogeneous magnetic films.
Three phases are identified: They are pinning phase when the disorder strength
is above a critical value that depends on the driving current density.
Below the critical disorder strength, the skyrmion transverse motion is boosted by the disorder
below a critical damping while the transverse motion is hindered above the critical damping.
The critical damping depends also on the current density and the disorder strength.
We showed that boosting of the transverse motion is mainly due to the random force opposite
to the current direction. We further demonstrated that the generalised Thiele equation can
perfectly capture skyrmion dynamics with a random force. Similar to the friction force in
Newtonian mechanics in which there exist static and kinetic friction forces, the random
force on a skyrmion can be classified as the static random force and kinetic random force.
For a pinned skyrmion, the static force is always transverse to the current direction
such that random force balances the current driving force $\vec G \times \vec u$, where
$\vec u$ is the usual quantity that characterise the Slonczewski spin-transfer torque.
When the skyrmion is in motion, the direction of the kinetic random force is opposite to
the skyrmion velocity, and the value of the kinetic random force is proportional to the
$(\Delta K)^2/u$.

\section{Acknowledgement}
This work is supported by by the NSFC Grant (No. 11974296 and 11774296)
and Hong Kong RGC Grants (No. 16301518, 16301619
and 16300117). HYY was financially supported by National Natural
Science Foundation of China (Grants No. 61704071).

\section{Appendix}
\begin{figure}
\centering
\includegraphics[width=0.9\columnwidth]{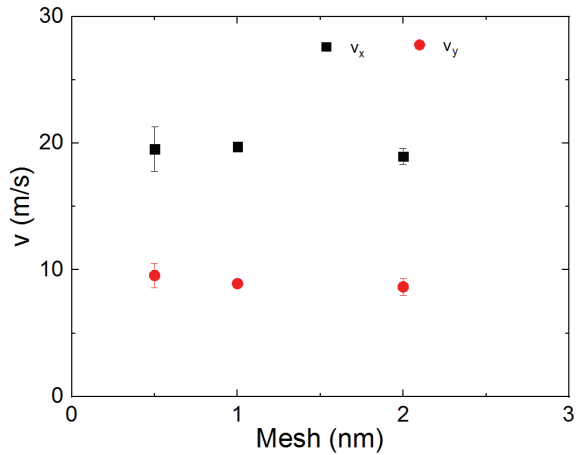}
\caption{(color online) The mesh size dependence of skyrmion velocities for $J =6\times 10^{11} \rm{A/m^2}$ and
$\Delta K = 3\%K_0,\alpha=0.3$.}
\label{fig10}
\end{figure}

In this appendix, we show further reduce mesh size
below 1 nm will not change simulation results within model parameters used. 
We simulate the skyrmion velocity using different mesh size,
as shown in Fig. \ref{fig10}. When mesh is larger than 4 nm, the skyrmion is unstable.
As mesh decreases from 2 nm to 0.5 nm, the skyrmion velocity does not change significantly. 
This indicates that 1 nm mesh used in the main text is sufficient.

\newpage
\end{document}